\documentclass{aa}
\usepackage{graphicx}
\usepackage{txfonts}
\begin{document}
   \title{Computation of the Fourier parameters of 
    RR~Lyrae stars by template fitting}

   \author{G. Kov\'acs\inst{1} \and G. Kupi\inst{2}}


   \institute{
   Konkoly Observatory, P.O. Box 67, H-1525,
   Budapest, Hungary \\ \email {kovacs@konkoly.hu}
   \\
   \and ARI, Heidelberg \\ \email {kupiga@ari.uni-heidelberg.de}
   }

   \date{July 16, 2006 / September 28, 2006}

%
%
   \titlerunning {RR~Lyrae template fitting }
   \abstract
{}
{Due to the importance of accurate Fourier parameters, we devise a 
method that is more appropriate for deriving these parameters on 
low-quality data than the traditional Fourier fitting.}
{Based on the accurate light curves of 248 fundamental mode RR~Lyrae 
stars, we test the power of a full-fetched implementation of the 
template method in the computation of the Fourier decomposition. 
The applicability of the method is demonstrated also on datasets 
of filter passbands different from that of the template set.} 
{We examine in more detail the question of the estimation of 
Fourier-based iron abundance [Fe/H] and average brightness. 
We get, for example, for light curves sampled randomly in 30 data 
points with $\sigma=0.03$~mag observational noise that optimized 
direct Fourier fits yield $\sigma({\rm [Fe/H]})=0.33$, whereas the 
template fits result in $\sigma({\rm [Fe/H]})=0.18$. Tests made on 
the RR~Lyrae database of the Large Magellanic Cloud (LMC) of the 
Optical Gravitational Lensing Experiment (OGLE) support the 
applicability of the method on real photometric time series. These 
tests also show that the dominant part of error in estimating the 
average brightness comes from other sources, most probably from 
crowding effects, even for under-sampled light curves.} 
{}


   \maketitle
%

%
%

\section{Introduction}
The method of Template Fitting (TF) is a widely used approach in 
data analysis. In astronomical applications we find examples from 
spectrum analysis (e.g., Bertone et al. 2004) to galaxy 
classification and redshift estimation (e.g., Wolf, Meisenheimer 
\& R\"oser 2001; Padmanabhan et al. 2005). The method is based 
on the simple assumption that the population to be studied contains 
targets sharing the same topological properties as the members 
of the template set. The latter is defined as a set containing 
all possible `flavors' of the given population and possessing 
accurately known parameter arrays -- e.g., spectrum, redshift, 
etc. The actual implementation of the TF method ranges from the 
simple few-template direct match (e.g., Jones, Carney \& Fulbright 
1996; Layden 1998) to the more sophisticated Artificial Neural 
Network method (e.g., Collister \& Lahav 2004).   

Here we present a `brute force' direct TF method that is aimed 
at the computation of the Fourier decompositions of fundamental 
mode RR~Lyrae (RRab) stars. The goal of this investigation is to 
provide a reliable method for the computation of the Fourier 
decomposition of {\it any} observed RRab light curve even if 
observational noise or poor sampling impair standard Fourier 
decomposition. Our approach is different from that of Kanbur \& 
Mariani (2004) and Tanvir et al. (2005), who employed principal 
component analysis (PCA) to parametrize the light curves of 
RR~Lyrae and Cepheid variables. With the aid of PCA one is able 
to create a smaller set from the templates and represent the 
target by a low-degree (in some sense optimum) PCA decomposition. 
However, low-order PCA decompositions are often insufficient 
if more subtle features are required to fit (see also Fig.~2 of 
Tanvir et al. 2005). In addition, in the case of RR~Lyrae stars, 
Blazhko effect further increases the possible types of light 
curves(see also Jurcsik, Benk\H o \& Szeidl, 2002). Therefore, 
we are resorted to a method that is able to handle a large variety 
of light curves, flexible enough but does not `over-fit' the data. 

Compared to earlier related works on RR~Lyrae stars, the present 
one utilizes a much larger template set, containing 248 accurate  
RRab light curves. The method is tested through a comparison with 
an optimized Fourier fit. We focus on the accuracy of the estimation 
of the iron abundance [Fe/H] based on the Fourier decomposition 
(Jurcsik \& Kov\'acs 1996, hereafter JK96, see also Kov\'acs 2005) 
and on the determination of the period-luminosity-color (PLC) 
relation (Kov\'acs \& Walker 2001, hereafter KW01). Neither of 
these quantities can be accurately estimated with direct Fourier 
fit if the number of data points is low and the noise is high, 
such as in the case of the $V$ and $B$ band observations of OGLE 
(Soszynski et al. 2003). Furthermore, the success of any utilization 
of the color indices in the computation of the physical parameters 
(most importantly that of $T_{\rm eff}$) strongly depends on the 
accuracy by which we estimate average colors. As far as the computation 
of the Fourier-based [Fe/H] is concerned, here the accurate estimation 
of $\varphi_{31}$ is also important, because of the strong 
dependence of [Fe/H] on this quantity -- see the application of 
the empirical [Fe/H] formula on the MACHO LMC data by Kunder et al. 
(2006). Current efforts in deriving [Fe/H] on large samples of 
stars in globular clusters and nearby galaxies from 
low/medium-dispersion spectroscopy (e.g., Sandstrom, 
Pilachowski \& Saha 2001 [M3]; Gratton et al. 2004 [LMC]; 
Clementini et al. 2005a [NGC 6441]; Clementini et al. 2005b 
[Sculptor dwarf spheroidal galaxy]; Sollima et al. 2006 
[$\omega$~Cen]) make the accurate computation of the Fourier 
decompositions of RRab stars even more interesting. 

In the subsequent sections we describe the method, optimize the 
template function, discuss template completeness, investigate 
the effect of choosing different filter passbands for the target 
and template sets and test the accuracy of [Fe/H] and average 
magnitudes derived from the TF method. Finally, we present 
results based on a limited dataset from the OGLE RRab database.

%
%

\section{The TF and the direct Fourier methods}
In order to make a meaningful (and fair) comparison between the 
Fourier parameters derived from the TF method and those obtained 
from a straightforward Fourier fit, we need to perform the latter 
in an `optimum way' (i.e., the way it would be done by a skilled 
data-analyst by checking the fit for different orders and avoiding 
-- if possible -- overshooting and appearance of strong wiggles 
for under-sampled light curves). First we describe our automated 
method for the Direct Fourier Fitting (DFF) and then give details 
on the Template Fourier Fitting (TFF\footnote{A Fortran'77 source 
code of TFF is available at http://www.konkoly.hu/staff/kovacs/tff.html}). 

In the case of DFF we tried to establish a set of criteria that 
results in `good looking' fits and can be applied automatically, 
without further human inspection. Because we employ standard 
{\em unweighted least squares} Fourier fits, the only parameter 
by which we can influence the quality of the fit is the {\em order} 
of the Fourier sum. By scanning the orders from $1$ to $10$, we 
choose the highest order at which the following criteria are 
satisfied:
\begin{itemize}
\item
The Fourier amplitudes are still `nearly' monotonically decreasing 
(i.e. $A_{i-1}>f_AA_i$, where $f_A$ is one of our `trial and error' 
parameters and is set equal to $1.2$, based on the inspection of 
the observed Fourier amplitudes of RRab stars -- see Fig.~1).
\item
The unbiased estimate of the fitting accuracy (the r.m.s. of the 
residuals between the fit and the data) is minimum.
\item
The total amplitude of the fitted curve is not greater than 
$f_TA_{tot}$, where $A_{tot}$ is the total amplitude of the target 
signal and $f_T$ is our second empirical parameter, and is also 
chosen to be equal to $1.2$.
\end{itemize}

%
%
   \begin{figure}[h]
   \centering
   \includegraphics[width=85mm]{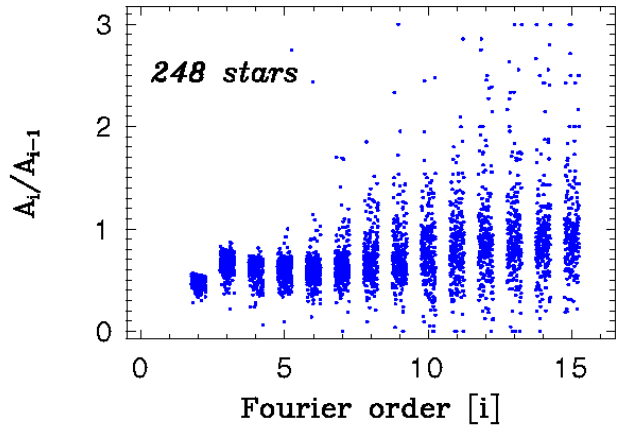}
      \caption{Amplitude ratios of the successive Fourier components 
               of the 248 RRab stars of the basic dataset. For better 
	       visibility, data points related to the individual objects 
	       are shifted in the horizontal direction. Except for 
	       orders above 9--10, for most of the objects the 
	       amplitudes monotonically decrease with the increase of 
	       the order of the components.} 
         \label{fig1}
   \end{figure}

These criteria make DFF reasonably stable for poorly sampled 
light curves, and produce accurate fit for well-sampled ones. 
We note that Ngeow et al. (2003) employed a somewhat similar 
method in deriving smooth and stable Fourier decompositions 
for Cepheid light curves. Their method employs 
`simulated annealing' (see Press et al. 1992) constrained 
by the period dependence of the Fourier amplitudes of Cepheids 
(the so-called Hertzsprung progression).   
 
For TFF, our approach is similar to that of Layden (1998), except 
that: (i) we use a much larger template set, based on individual 
variables and not on a limited set of visually selected classes; 
(ii) we allow low-degree polynomial transformation of the 
template in finding the best fit. 
 
First we choose a set of Fourier decompositions derived from 
well-observed, densely sampled light curves. Such a set is 
available from our earlier studies on the $V$ band light curves 
of RRab stars (see the CDS archive of KW01). There are altogether 
492 variables, with 105 stars from the Galactic field and the rest 
from various globular clusters and from the Sculptor dwarf galaxy. 
Because some of variables are poorly sampled, we apply the following 
selection criterion in order to employ only the best quality light 
curves. First we define the Quality Factor (QF) of a given light 
curve as a quantity proportional to the ratio of the amplitude $A_1$ 
of the first Fourier component to its simple error estimate, i.e. 
%
%
\begin{eqnarray}
{\rm QF} & = & \sqrt{N}A_1/\sigma_{\rm fit} \hskip 2mm ,
\end{eqnarray}
where $N$ is the number of data points and $\sigma_{\rm fit}$ is 
the unbiased estimate of the standard deviation of the residuals 
of the fit. The quality control is established by requiring 
$\rm{QF}>{\rm QF_{\rm min}}$, where ${\rm {QF}_{\rm min}}$ is a 
preset threshold. By changing ${\rm QF_{\rm min}}$ from $100$ to 
$200$, the number of the remaining stars  decreases from 336 to 193. 
Finally we decided to apply ${\rm QF_{\rm min}}=150$ and obtained 
a set of $251$ stars. The overwhelming majority of these stars have 
$N>100$, and there are only three stars with $N<40$. 
Since at lower number of data points there is a greater risk of 
having erroneous Fourier decompositions, we omit these three stars 
(\object{FH Vul}, \object{IV Hya} and \object{NGC1841 V4}). Finally 
we arrive to our {\em basic dataset} containing $248$ variables. 
We use this set as the template set throughout this paper.    

Once the template set is selected, for a target light curve we 
find the best fitting template in the following way.
\begin{itemize}
\item
Compute densely sampled folded light curves from the template 
Fourier decompositions. We denote these functions by 
\{$x_{i,j}(\varphi)$\}, where subscript $i$ stands for the array 
index in the folded light curve and $j$ refers to the template 
identification. Phase $\varphi$ of the template is arbitrary 
at this step.     
\item
Compute folded light curve \{$Y_i$\} for the target. 
\item
For each initial phase $\varphi$ and for each template, minimize 
the following quantity:
%
%
\begin{eqnarray}
{\cal D}_j(\varphi) & = & {1\over N}\sum_{i=1}^N 
[Y_i-X_i(\varphi)]^2 \hskip 2mm ,
\end{eqnarray}
where 
%
%
\[ X_i(\varphi) = \left\{
\begin{array}{ll}
c_0 + x_{i,j}(\varphi)              & \hskip 8mm \mbox{if \ $M=0$} \\
\sum_{k=0}^M c_k x_{i,j}^k(\varphi) & \hskip 8mm \mbox{if \ $M\gid 1$} 
\end{array}
\right. \]

Here $M$ denotes a preset polynomial degree, to be determined 
in Sect.~3 as a data quality-dependent parameter. While scanning 
$\varphi$, we employ quadratic interpolation for the template in 
order to get a good approximation for its value at the moments 
where the target is given.
\item
TFF is computed by the Fourier decomposition of that 
\{$X_i(\varphi)$\}, which minimizes ${\cal D}_j(\varphi)$ 
of Eq.~(2).
\item
The above steps are to be supplemented by the ones to be discussed 
in Sect.~3 for the optimum choice of $M$.
\end{itemize}
Typically we use template light curves sampled in $300$ points 
and require an accuracy of $10^{-5}$ in the phase match between 
the target and template. The search for optimum phase is made 
iteratively, starting with some 50 phase steps. Execution time 
is not an issue with current several GHz machines. 

One can construct other types of TF methods, by using different  
functional dependence of the target on the template members 
(e.g., linear or polynomial multi-template functions). However, 
in our approach we consider the current template set only as 
a {\em subset} of an ever growing {\em master set} that will 
be accumulated in the future. For this ideal set we might need 
only a scaling factor for a very precise fit of {\em any} target, 
because the master set will contain all `flavors' of RRab stars 
and the fitting routine needs to perform the search only among the 
single template members. In addition, more complicated functional 
dependence would make the fitting procedure slower.

%
%

\section{Template polynomial degree and completeness}
Before testing the TFF method described in Sect.~2, it is 
necessary to determine the best polynomial degree $M$ to 
be used in the template transformation. Furthermore, it 
is also important to examine if the template set with the 
adopted TFF method is capable of reproducing all (or most of) 
the light curve `flavors' observed among RRab stars. This 
latter property is connected to what we call `completeness' 
and to be defined somewhat more precisely later in this 
section. 

The purpose of introducing the polynomial template 
transformation is to increase our freedom in reaching 
higher accuracy in fitting targets. Obviously, employing 
a too high-degree polynomial may lead to instability, similarly 
as if we used high-order Fourier fit. Because the prime goal 
of the application of the template method is just to avoid 
this type of instability, we accept the lowest polynomial 
degree that yields fits of similar quality as the higher 
degree ones.      

In order to rank the results obtained with various polynomial 
degrees, we need to define a function that characterizes the 
quality of the fit for an ensemble of targets. Using some average 
of the standard deviations of the fits to the individual targets 
is not satisfactory, because then, poorly fitted small-amplitude 
variables may stay hidden due to the small standard deviations 
associated with their small amplitudes. A possible normalization 
by the amplitude (see Eq.~(1) for QF) may partially cure this 
problem, but we found more satisfactory to use a function that 
is independent of the amplitude and more closely related to our 
prime interest in deriving accurate phases. We introduce the 
following quantity to characterize the goodness of the fit 
%
%
\begin{eqnarray}
{\rm RMS}(\Delta\varphi) & = & 
\biggl[{1\over 3}\sum_{k=2}^4 (\Delta \varphi_{k1})^2\biggr]^{1\over 2} 
\hskip 2mm ,
\end{eqnarray}
where $\Delta \varphi_{k1}$ is the difference between the target 
and the best template-fitted phases. The epoch-independent phase 
is defined in the usual way $\varphi_{k1}=\varphi_{k}-k\varphi_{1}$ 
(Simon \& Lee 1981). Notice that the above expression utilizes 
all three low-order phases, not only $\varphi_{31}$ that enters 
in the empirical formula for [Fe/H]. We think that using more phases 
makes the results more stable against statistical fluctuations. 
At the same time, adding high-order phases would make the above 
statistic biased toward small fitting errors, that we would like 
to avoid.  

By using ${\rm RMS}(\Delta\varphi)$, the optimum polynomial 
degree $M$ is determined in the following way. For any fixed 
$M$ we start with the $248$ stars of the basic dataset and 
for each variable of this set we find the best fitting 
template selected from the remaining $247$ stars. The time 
series of the target is computed from the Fourier series 
given in the basic dataset for the target chosen. The time 
base of the sampling is taken arbitrarily as $12.3456P$, 
where $P$ is the period of the target. The sampling rate 
is quasi-uniform with a small randomness of the size of the 
exact uniform sampling. Tests are run both with and without 
noise added to the synthetic data.   
We note that other choices of the time base would also serve 
the purpose, except for near integer multiple of the period, 
when the chance of regular sampling of the phased light curve 
would be greater. 

Once the best template (i.e., \{$X_i(\varphi)$\}, in Eq.~(2)) 
is found, the resulting Fourier decomposition is compared with 
that of the target via Eq.~(3). In this way at each fixed 
$M$ we get $248$ ${\rm RMS}(\Delta\varphi)$ values that can 
be analyzed statistically and compare with those obtained at 
other $M$ values. 

The most straightforward way to compare the 
${\rm RMS}(\Delta\varphi)$ values is to compute their 
probability distribution function (PDF). In Fig.~2 we 
show such functions computed for the noiseless (i.e., 
$\sigma=0.0$) simulations with $N=20$. As expected, the 
distribution functions become more narrow and concentrated 
at lower ${\rm RMS}(\Delta\varphi)$ values when $M$ is chosen 
to be low. From this test the optimum $M$ at $N=20$ is expected 
between $2$ and $4$ for noiseless data.  

%
%
   \begin{figure}[h]
   \centering
   \includegraphics[width=85mm]{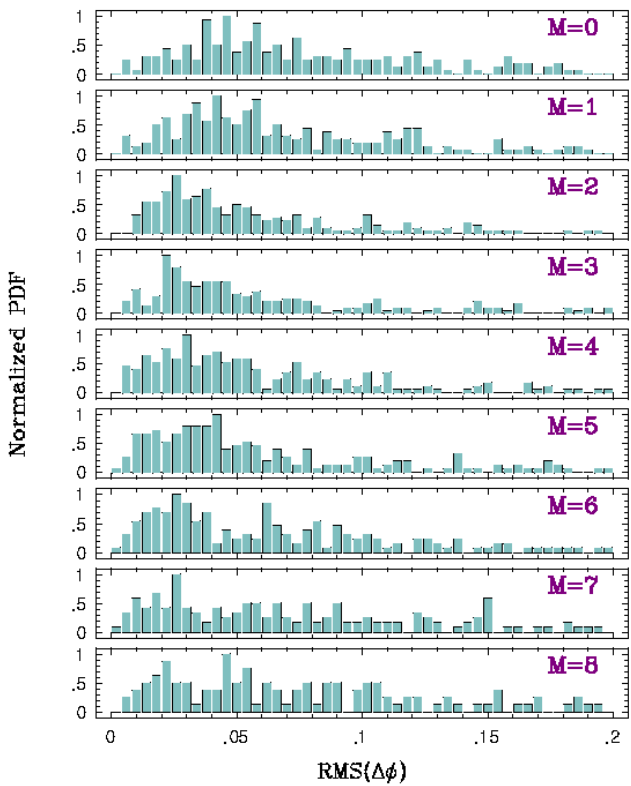}
      \caption{Probability distribution functions of the 
               average phase differences ${\rm RMS}(\Delta\varphi)$ 
	       (Eq.~3) for test targets at various template 
	       polynomial degrees $M$. All target light curves were 
	       generated with $N=20$ data points. No noise is added. 
	       The test was performed on the basic dataset. Each PDF 
	       is normalized to its maximum. We draw attention to the 
	       narrowing of the PDFs at low $M$ values of $2$--$4$.} 
         \label{fig2}
   \end{figure}

If we perform the above test at other values of $N$, 
then we get different values for the optimum $M$. 
This is understandable, because at low $N$ the high 
polynomial degree leads to stronger instabilities, 
whereas at high $N$ we are able to fit templates of 
high degree, thereby reaching higher accuracy. 
In order to get an estimate on the size of this shift 
of $M$, we performed additional tests with $N=15$, 
$N=40$ and $N=100$. For an easier comparison, instead of 
plotting PDFs, we compute the number of stars 
that have ${\rm RMS}(\Delta\varphi)$ greater than 
${\rm RMS}_{\rm max}$, where the latter quantity is chosen 
in a way which ensures that, at least for some distribution 
functions, the number of stars satisfying this condition is 
small. From the inspection of Fig.~2 we choose 
${\rm RMS}_{\rm max}=0.12$, because for weakly-spread 
distribution functions the tail seems to be separated 
from the bulk of the distribution at this value. (We note 
that our conclusion does not change by choosing other 
cutoff values in the range of $0.07$--$0.17$.)

%
%
   \begin{figure}[h]
   \centering
   \includegraphics[width=85mm]{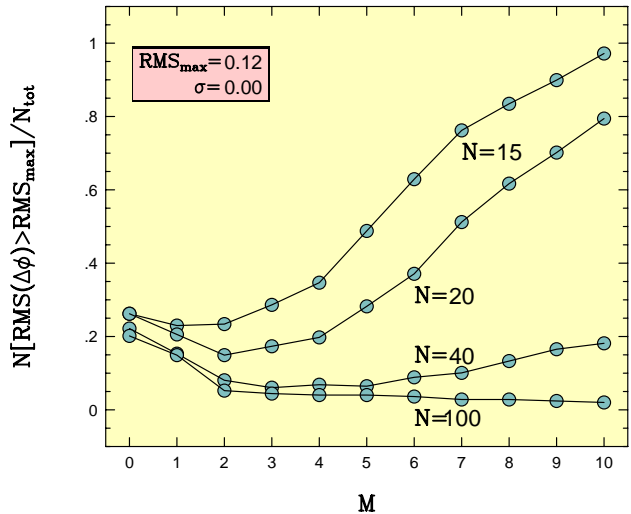}
      \caption{Dependence of the relative number of poorly fitted targets 
               on the polynomial template degree $M$ and on the number of 
	       data points $N$ of the target time series. Tests were made 
	       on the basic dataset with $N_{\rm tot}=248$ stars. No noise 
	       was added to the synthetic data.
	       } 
         \label{fig3}
   \end{figure}
%
%

%
%
   \begin{figure}[h]
   \centering
   \includegraphics[width=85mm]{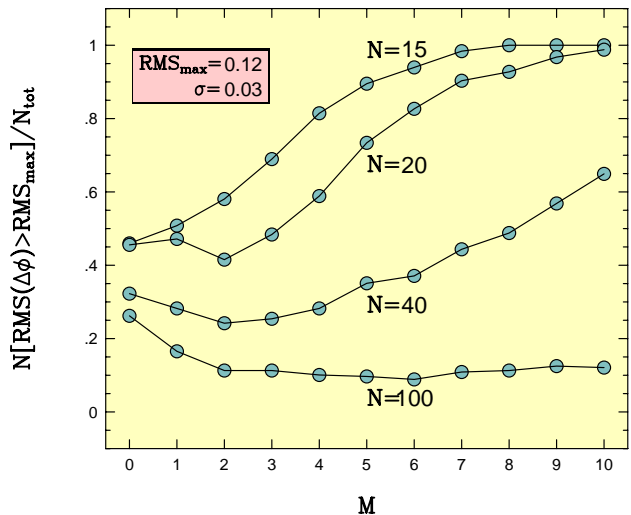}
      \caption{As in Fig.~3, but with added noise as given in the 
               shaded box.} 
         \label{fig4}
   \end{figure}

Fig.~3 shows that the size of the shift in the location 
of the minima of the functions is small for low $N$ values 
and we can still stay in $M=2$--$4$, with hitting the low 
and high boundaries at low and high $N$ values, respectively. 
For modest (or high) $N$ values the minima become shallower, 
therefore, there will not be much difference between 
choosing moderately low $M$ and the optimum one. However, 
it is clear that at large $N$ the minimum will be lower 
and gradually shifted to rather high $M$ values. On the 
other hand, these high-$N$, well-sampled cases are not 
interesting from the present point of view, because these 
are well-treated by standard Fourier fitting methods. 
(Exceptions are of course cases when high noise prohibits 
the traditional approach, and we will become better off 
again by using the template method at low $M$.) 

To check the effect of noise on the optimum polynomial 
degree, we repeated the above test by adding moderate 
Gaussian noise of $\sigma =0.03$~mag to the synthetic 
target signals. The result is shown in Fig.~4. As 
expected, the properties observed at low $N$ in the 
noiseless case of Fig.~3 are shifted to higher $N$. 
At the same time, the low-$N$ signals get basically 
out of control at high $M$. 

Due to the dependence of the optimum template polynomial 
degree on data quality, we need to examine this relation 
more closely. The data quality is characterized by the 
Signal-to-Noise Ratio (SNR), defined similarly to QF
%
%
\begin{eqnarray}
{\rm SNR} & = & \sqrt{N}A/\sigma_{\rm fit} \hskip 2mm ,
\end{eqnarray}
where $A$ is the {\em total} amplitude of the light variation, 
$\sigma_{\rm fit}$ is the unbiased estimate of the standard 
deviation of the residuals between the target and the fit. 
We performed tests similar to the ones described above and 
examined the behavior of ${\rm RMS}(\Delta\varphi)$ as a 
function of SNR and $M$. The SNR values were obtained from 
the results at $M=1$. For $N$, $M$ and $\sigma$ we took the 
following values: $N=15,\ 30,\ 45,\ 60,\ 90$, $M=0,\ 1,\ 2,\ 3,\ 4$ 
and $\sigma=0.0,\ 0.03,\ 0.06,\ 0.09$. Simulations corresponding 
to the same $M$ were averaged and plotted as functions of 
SNR in Fig.~5. The errors of the averages are about the size 
of the circles, except toward the low SNR end, where they  
increase to $0.1$--$0.2$. It is clear from this figure that 
the optimum polynomial degree is a function of SNR. Therefore, 
the TFF algorithm described in Sect.~2 is supplemented by the 
following steps in selecting the optimum polynomial degree.
\begin{itemize}
\item
Compute SNR by applying TFF at $M=1$
\item
Choose the best $M$ depending on the computed SNR
%
%
\[ M = \left\{
\begin{array}{ll}
0,   & \hskip 5mm \mbox{if \ ${\rm SNR}<50.0 $} \\
1,   & \hskip 5mm \mbox{if \ $50.0\lid{\rm SNR}\lid 150.0 $} \\
2,   & \hskip 5mm \mbox{if \ ${\rm SNR}>150.0$} 
\end{array}
\right. \]
\item
Compute TFF with $M$ determined above
\end{itemize}
We note that SNR can also be estimated by the simplest $M=0$ fit, 
but we choose $M=1$, because it yields a somewhat more consistent 
result (i.e., cleaner separation of the averages corresponding to 
the various $M$ values) and because of the unavoidable scaling in 
fitting light curves of different passbands (see Sect.~5). In the 
rest of this paper we use TFF with the above optimized $M$.

%
%
   \begin{figure}[h]
   \centering
   \includegraphics[width=85mm]{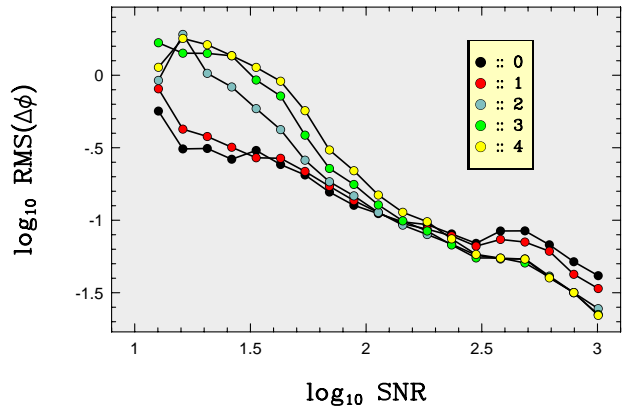}
      \caption{Dependence of the average of ${\rm RMS}(\Delta\varphi)$ 
               on SNR at various template polynomial degree $M$ shown 
	       in the inset. The figure is based on simulated data 
	       described in the text in detail.} 
         \label{fig5}
   \end{figure}

Next we address the question of template completeness. 
As already mentioned, we would like to measure the ability 
of the basic dataset in reproducing each member of the 
set with the aid of the TFF method. It is clear that we 
need to define the meaning of the word `reproduce'. Here, 
employing the same argument as earlier, we use 
${\rm RMS}(\Delta\varphi)$ as a quantity characterizing 
the goodness of the fit. Then we say that the template 
set is complete at the level of  ${\rm C}_{\rm TFF}$ 
at ${\rm RMS}_{\rm max}$, if the fraction of variables 
that satisfy the condition 
${\rm RMS}(\Delta\varphi)<{\rm RMS}_{\rm max}$ is equal to 
${\rm C}_{\rm TFF}$. Because the introduction of 
${\rm C}_{\rm TFF}$ is aimed at to characterize the ability 
of the template set to `reproduce' itself by using TFF, we 
generate noiseless, well-sampled, high-$N$ light curves to 
derive the dataset necessary for the computation of 
${\rm C}_{\rm TFF}$. The result of is shown in Fig.~6. 
We see that the completeness is close to 90\% at 
${\rm RMS}_{\rm max}=0.1$ and reaches 95\% at  
${\rm RMS}_{\rm max}=0.15$. For some 80\% of the stars we 
get matches with smaller RMS than $0.05$.  

%
%
   \begin{figure}[h]
   \centering
   \includegraphics[width=85mm]{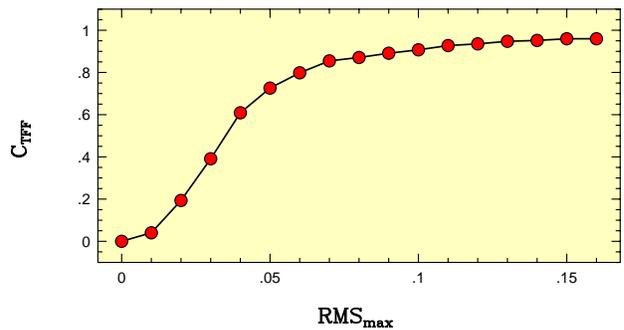}
      \caption{Completeness function of the TFF method. The symbol 
               ${\rm C}_{\rm TFF}$ denotes the relative number of 
	       variables that have 
	       ${\rm RMS}(\Delta\varphi)<{\rm RMS}_{\rm max}$. 
	       We used the 248 light curves of the basic dataset 
	       with $N=200$ synthetic data points per object and 
	       polynomial degree $M=2$.} 
         \label{fig6}
   \end{figure}

Although the above numbers indicate a reasonable completeness, 
we note that there are stars that stubbornly resist to accurate 
template fitting and prevent high completeness even at low-accuracy 
(e.g. for ${\rm RMS}_{\rm max}>0.15$). For example, variable 
\object{M107~V12}, cannot be fitted, yielding 
${\rm RMS}(\Delta\varphi)=0.263$ and 
$\Delta\varphi_{31}=\varphi_{31}({\rm target})-\varphi_{31}({\rm template})=0.389$. 
There are some $15$ stars that have $\vert\Delta\varphi_{31}\vert>0.1$. 
The fact that some stars cannot be fitted even if their light curves 
are very densely sampled, can be explained by one (or some) of the 
following reasons: (i) the size of the current template set is small, 
and therefore, it is unable to reproduce some of the existing 
light curves with a desirable accuracy; (ii) some of the stars 
may exhibit long-periodic amplitude- and phase-modulations that 
result in discrepant template matches; (iii) the data on the targets 
were insufficient (too few data points, gaps in the folded light 
curves, etc.), that has led to inaccurate Fourier decompositions 
(in spite of these stars passing through our criteria for template 
membership); (iv) instrumental effects (e.g. daily or seasonal drifts) 
make the given light curve unique and therefore, not treatable by 
the template method; (v) some stars have periods close to integer 
ratios of one day, that again, may lead to unique light curves, 
especially if it is combined with property (iv). We think that 
from the present dataset we cannot decide which of the above effects 
is responsible for the outlier status of some of the stars (a closer 
examination of some of the outliers has shown that one can find 
examples/suspects for all these five possibilities). Considering 
that the current template set contains several stars with limited 
coverage either in time or in amount of data points, the existence 
of the few outliers is not surprising. It is clear that the present 
template set should be further extended by utilizing new observations 
and by leaving out objects with poor quality light curves.     

To illustrate the difference between the matches produced by 
DFF and TFF at low number of data points, in Fig.~7 we show two 
examples. Although for other data distributions DFF might behave 
less erratically, the behavior shown is quite common at low number 
of data points. We note that in many cases the best matching 
template may have widely different period from that of the target. 
For instance, in the example shown, \object{DX Del} has a period 
of $0.47262$~d, whereas the best match is produced by 
\object{$\omega$~Cen V176}, that has a period of $0.74275$~d. 
For \object{RR Leo} the situation is different, here we have 
$P=0.45239$~d for the target and  $P=0.45930$~d for the matching 
template \object{SW Aqr}. 

%
%
   \begin{figure}[h]
   \centering
   \includegraphics[width=85mm]{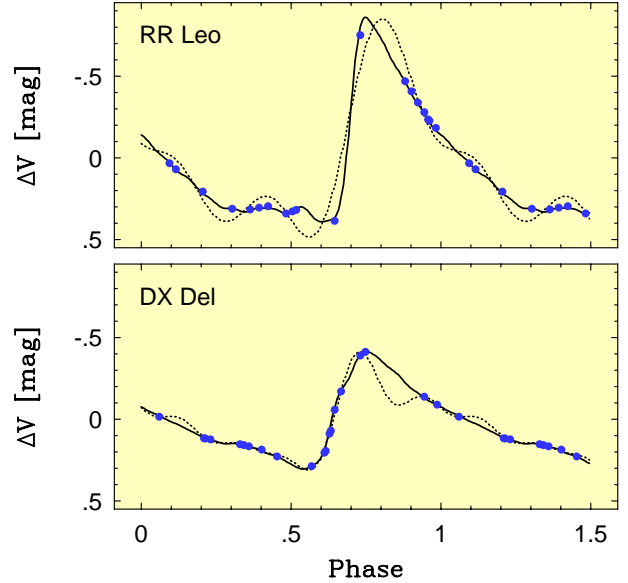}
      \caption{Examples of the performance of TFF at low data point 
               numbers. Dots, dashed and continuous lines are for the 
	       noiseless synthetic data, DFF and TFF fits, respectively. 
	       The noiseless synthetic data have been obtained by 
	       the sparse sampling of the accurate observed light curve.
	       } 
         \label{fig7}
   \end{figure}
%

%
%

\section{Estimation of the photometric [Fe/H]}
From the point of view of applications, it is important to examine 
the accuracy of the TFF method in determining the photometric [Fe/H] 
with the aid of the formula of JK96. We used the basic dataset in 
generating simulated light curves with the following parameter values: 
$N=15,\ 30,\ 45,\ 60,\ 90$; $\sigma=0.0,\ 0.03,\ 0.06$. The polynomial 
degree $M$ was optimized as described in Sect.~3. For each $N$, 
$\sigma$ and for each star, we compute $\Delta{\rm [Fe/H]}$, the 
difference between the [Fe/H] values computed from the target and 
from the fitted light curves. From these values we derive: 
$\sigma(\Delta{\rm [Fe/H])}$, the standard deviation of the [Fe/H] 
differences; $R01$, the ratio of the stars with 
$|\Delta{\rm [Fe/H]|<0.1}$ to the total number of stars (i.e. to $248$); 
$T/D$, the ratio of the number of stars for which $|\Delta{\rm [Fe/H]|}$ 
is smaller for TFF than for DFF to the ones for which the opposite is 
true. 

%
%
\begin{table}[h]                                                     
\caption[]{Accuracy of the determination of [Fe/H].}                                                    
\begin{flushleft}                                                     
\begin{tabular}{ccccccc}                                         
\hline\hline
$\sigma$ & $N$ & $\sigma_{\rm D}$ & $\sigma_{\rm T}$ & 
$R01_{\rm D}$ & $R01_{\rm T}$ & $T/D$ \\
\hline
0.00 &    15 &  0.66 &  0.15 &   0.2 &   0.7 &   4.4 \\
0.00 &    30 &  0.21 &  0.11 &   0.7 &   0.9 &   1.5 \\
0.00 &    45 &  0.12 &  0.09 &   0.9 &   0.9 &   0.6 \\
0.00 &    60 &  0.03 &  0.09 &   1.0 &   0.9 &   0.3 \\
0.00 &    90 &  0.02 &  0.08 &   1.0 &   0.9 &   0.2 \\
\hline
0.03 &    15 &  0.76 &  0.27 &   0.2 &   0.4 &   3.9 \\
0.03 &    30 &  0.33 &  0.18 &   0.4 &   0.5 &   1.9 \\
0.03 &    45 &  0.24 &  0.17 &   0.5 &   0.6 &   1.4 \\
0.03 &    60 &  0.13 &  0.15 &   0.7 &   0.7 &   1.0 \\
0.03 &    90 &  0.12 &  0.13 &   0.7 &   0.7 &   0.8 \\
\hline
0.06 &    15 &  0.86 &  0.37 &   0.1 &   0.3 &   4.9 \\
0.06 &    30 &  0.45 &  0.29 &   0.3 &   0.3 &   1.6 \\
0.06 &    45 &  0.36 &  0.22 &   0.3 &   0.4 &   1.9 \\
0.06 &    60 &  0.24 &  0.19 &   0.4 &   0.5 &   1.8 \\
0.06 &    90 &  0.23 &  0.18 &   0.4 &   0.5 &   1.5 \\
\hline                                                                
\end{tabular}                                                         
\end{flushleft}                                                       
{\footnotesize\underline {Note:} 
$\sigma$: standard deviation of the noise added to the synthetic 
light curves that were generated from the template Fourier 
decompositions; 
$N$: number of data points; 
$\sigma_{\rm D}$: standard deviation of 
$\Delta {\rm [Fe/H]}_{\rm D}\equiv
{\rm [Fe/H]}_{\rm target}-{\rm [Fe/H]}_{\rm DFF}$;  
$\sigma_{\rm T}$: as $\sigma_{\rm D}$, but for TFF;
$R01_{\rm D}$: number of stars with $|\Delta {\rm [Fe/H]}_{\rm D}|<0.1$ 
divided by the total number of stars of $248$; 
$R01_{\rm T}$: as $R01_{\rm D}$ but for TFF;  
$T/D$: number of stars with 
$|\Delta {\rm [Fe/H]}_{\rm T}|<|\Delta {\rm [Fe/H]}_{\rm D}|$  
divided by the number of stars satisfying the opposite inequality. 
The result is based on the $248$ stars of the basic dataset. 
}                                          
\end{table}

In Table~1 we show the average values obtained for these quantities 
for the two methods. The following conclusions can be drawn from this 
table:
\begin{itemize}
\item
Except at low noise level and high (i.e. $\ga 45$) data point number, 
TFF always yields more accurate [Fe/H] values in the average sense.   
\item
Within the above limit, the number of accurate [Fe/H] estimates 
(i.e. those with $|\Delta{\rm [Fe/H]|<0.1}$) is always larger for TFF.
\item
Within the above limit, the number of cases when TFF yields smaller 
error than DFF is always larger than that of the opposite situation. 
\item
This better performance is especially well visible at higher noise 
levels. 
\end{itemize} 

%
%
   \begin{figure}[h]
   \centering
   \includegraphics[width=85mm]{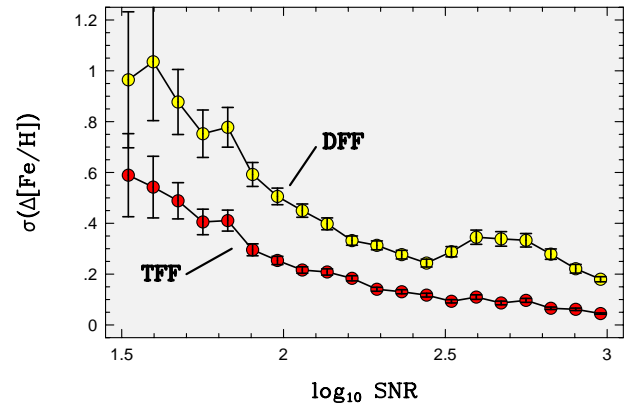}
      \caption{Standard deviation of the difference between the target 
               and computed Fourier [Fe/H] values as a function of 
	       SNR. The test was made on the $248$ stars of the basic 
	       dataset. Error bars show the $\pm1\sigma$ ranges of the 
	       means.} 
         \label{fig8}
   \end{figure}

To make a comparison yet in another parameter domain, in Fig.~8 
we plot the standard deviation of $\Delta{\rm [Fe/H]}$ as a 
function of SNR. Because the various simulations yield different 
individual SNR values, the total range of SNR was divided into 
$20$ bins and the standard deviations of the various 
$\Delta{\rm [Fe/H]}$ values within these bins have been computed. 
Except for low SNR values (i.e. for $\log{\rm SNR}<1.75$), all 
bins contain some 100--300 simulations (at $\log{\rm SNR}=1.5$ 
we have only 10). Although the scatter within the bins is very 
large, the averages are fairly accurately estimated and the 
difference between the two methods is clearly visible. This figure 
may give some guidance to a rough error estimation of the methods. 
In general, we may expect rather large errors -- in the average sense 
-- if $\log{\rm SNR}<2.0$. The average errors can be substantially 
decreased for DFF for $\log{\rm SNR}>2.0$ if we employ $3\sigma$ 
clipping on the derived [Fe/H] values. In this way the two methods 
will perform nearly in the same way for $\log{\rm SNR}>2.0$. The 
effect of $3\sigma$ clipping on TFF is minimal, because the number 
of outliers is much lower when TFF is employed. As an example, at 
$\log{\rm SNR}=2.75$, with $3\sigma$ clipping we loose $23$\% of 
the stars in the case of the DFF method. The same figure for TFF 
is only $2$\%.

%
%

\section{Estimation of the average magnitudes in various colors}
We test the applicability of the TFF method in estimating the 
average magnitudes of light curves observed in various wavebands. 
Unfortunately, the amount of good-quality multicolor data available 
for us is much lower than that of the single-color data. For 
simplicity, we take a set from the globular cluster data used by 
KW01 for the derivation of the PLC relation for RRab stars. The 
present set contains B, V and I light curves from globular clusters 
NGC~1851, 4499, 6362 and 6981. There are 95, 95  and 83 variables 
in B, V and in I colors, respectively. Although the amount and the 
quality of these data are less favorable than the ones used for 
testing the accuracy of the photometric [Fe/H], they are sufficient 
to get rough error estimates on the determination of average 
magnitudes by using different methods.     

%
%
   \begin{figure}[h]
   \centering
   \includegraphics[width=85mm]{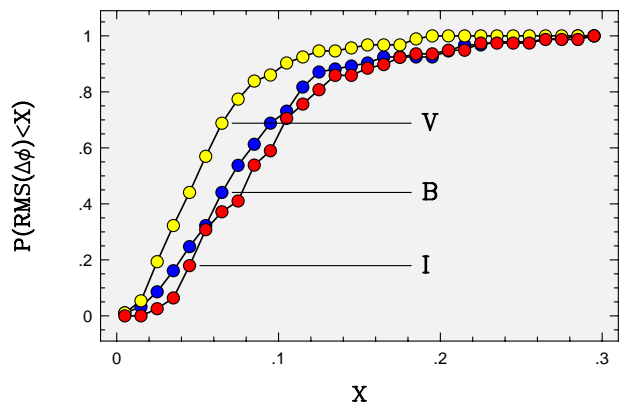}
      \caption{Empirical probability distribution functions 
               of ${\rm RMS}(\Delta\varphi)$ for noiseless test 
	       signals drawn from samples of RRab stars of different 
	       wavebands. All test signals have $N=90$ data points. 
	       The TFF method with the basic dataset (available only 
	       in color V) is used in the computation of 
	       $\Delta\varphi$.} 
         \label{fig9}
   \end{figure}
%

%
%
\begin{table}[h]                                                     
\caption[]{Accuracy of the magnitude averages.}                                                    
\begin{flushleft}                                                     
\begin{tabular}{ccccccc}                                         
\hline\hline
Band & AV/N & $15$ & $30$ & $45$ & $60$ & $90$ \\
\hline
      &        &      &  $\sigma=0.0$    &      &      &      \\
\hline
B:    &   AVE: &   0.082 & 0.051 & 0.032 & 0.027 & 0.017\\
      &   DFF: &   0.037 & 0.009 & 0.004 & 0.002 & 0.001\\
      &   TFF: &   0.030 & 0.007 & 0.005 & 0.003 & 0.002\\
V:    &   AVE: &   0.063 & 0.040 & 0.025 & 0.021 & 0.013\\
      &   DFF: &   0.032 & 0.007 & 0.004 & 0.002 & 0.001\\
      &   TFF: &   0.010 & 0.005 & 0.003 & 0.002 & 0.002\\
I:    &   AVE: &   0.039 & 0.023 & 0.013 & 0.013 & 0.008\\
      &   DFF: &   0.023 & 0.004 & 0.003 & 0.002 & 0.001\\
      &   TFF: &   0.012 & 0.004 & 0.002 & 0.001 & 0.001\\
\hline
      &        &      &  $\sigma=0.03$    &      &      &      \\
\hline                                                                
B:    &   AVE: &   0.082 & 0.051 & 0.032 & 0.027 & 0.018\\
      &   DFF: &   0.043 & 0.011 & 0.007 & 0.005 & 0.004\\
      &   TFF: &   0.027 & 0.008 & 0.007 & 0.005 & 0.004\\
V:    &   AVE: &   0.064 & 0.040 & 0.024 & 0.022 & 0.014\\
      &   DFF: &   0.030 & 0.014 & 0.006 & 0.005 & 0.003\\
      &   TFF: &   0.022 & 0.007 & 0.006 & 0.004 & 0.003\\
I:    &   AVE: &   0.040 & 0.024 & 0.013 & 0.014 & 0.008\\
      &   DFF: &   0.024 & 0.008 & 0.006 & 0.005 & 0.004\\
      &   TFF: &   0.015 & 0.007 & 0.005 & 0.004 & 0.003\\
\hline
      &        &      &  $\sigma=0.06$    &      &      &      \\
\hline
B:    &   AVE: &   0.083 & 0.052 & 0.032 & 0.028 & 0.019\\
      &   DFF: &   0.045 & 0.020 & 0.011 & 0.009 & 0.007\\
      &   TFF: &   0.032 & 0.013 & 0.011 & 0.008 & 0.007\\
V:    &   AVE: &   0.065 & 0.041 & 0.025 & 0.023 & 0.015\\
      &   DFF: &   0.037 & 0.018 & 0.011 & 0.008 & 0.007\\
      &   TFF: &   0.024 & 0.012 & 0.010 & 0.008 & 0.006\\
I:    &   AVE: &   0.042 & 0.027 & 0.014 & 0.015 & 0.010\\
      &   DFF: &   0.029 & 0.015 & 0.010 & 0.008 & 0.007\\
      &   TFF: &   0.029 & 0.013 & 0.009 & 0.008 & 0.006\\
\hline
\end{tabular}                                                         
\end{flushleft}                                                       
{\footnotesize\underline {Note:}
Each item of data corresponds to the standard deviation of the 
$\Delta A_0$ values, as given by the difference between the 
average magnitude of the target and the one obtained by the 
application of the various methods (AVE denotes the simple 
arithmetic average). The tests are based on cluster RRab stars 
as given in the text. The result depends somewhat on the 
realization used, but this does not change the basic trends 
shown here. 
}                                          
\end{table}

Before discussing results on the average magnitudes, it is 
interesting to compare the accuracy of the fit in the different 
colors. As before, we use ${\rm RMS}(\Delta\varphi)$ to characterize 
the goodness of fit. Figure~9 shows the PDFs of this quantity for 
the different colors. As expected, the V-band data are fitted 
most accurately. The B and I data perform similarly, with a slight 
preference (but probably within the error limits of the present 
test) toward the B data. The fact that these, seemingly very 
similar light curves are not transformable by TFF, shows that they 
contain independent pieces of information on the pulsation. In the 
present context this result suggests to avoid the use of templates 
of different waveband from that of the target if we are aimed at 
the estimation of the Fourier decomposition with the aid of the 
TFF method. 

In comparing the average magnitudes, we proceed in the same way as 
in the test of [Fe/H] in Sect.~4, except that now the target sets 
are limited on the cluster data mentioned above. The results are 
ranked on the basis of the standard deviations computed from the 
differences between the averages of the target (as given by the 
zero frequency constant in its accurate Fourier decomposition) and 
the estimated values. Table~2 shows the result of the computation 
for the three colors.   

Although other realizations (determining the data distribution and 
noise) lead to somewhat different results, the following basic trends 
seen in the table survive.
\begin{itemize}
\item
The simple arithmetic average (AVE) has the largest scatter and it 
is almost independent of the noise level.
\item
Within the statistical limits, TFF {\em always} performs better than 
DFF. 
\item
If the number of data points is greater than $\sim 40$, TFF and DFF 
yield results of similar accuracy. 
\end{itemize} 
The near independence of AVE on the noise level is due to the fact 
that in the course of simple averaging the main source of error is 
the uneven distribution of data points in the phased light curve. 
It is seen that this effect is not negligible even at high number 
of data points. Therefore, for the accurate computation of the 
averages, simple averaging should {\em never} be used. Although 
Fourier decompositions are less accurately estimated for light curves 
of colors different from that of the template set, averages have 
nearly the same accuracy in all colors.

%
%

\section{PLC estimates on artificial data}
The correlation of the period and the reddening-free magnitude 
$W_{B-V}=V-R_V(B-V)$ is an important relation in estimating 
RR~Lyrae distances from colors in the visual wavebands 
(Dickens \& Saunders 1965; Kov\'acs \& Jurcsik 1997; KW01; 
see also Kov\'acs 2003 for application in Baade-Wesselink 
analysis and Di~Criscienzo, Marconi \& Caputo 2004 for 
theoretical interpretation). Because of the large value of the 
selective absorption coefficient $R_V$, random errors in the 
estimated mean magnitudes become amplified in $W_{B-V}$ and 
thereby impair the accuracy by which we can employ the 
relation, among others, for distance determination. Therefore, 
it is worthwhile to test the effect of TFF in decreasing the 
error of $W_{B-V}$. 

%
%
   \begin{figure}[h]
   \centering
   \includegraphics[width=85mm]{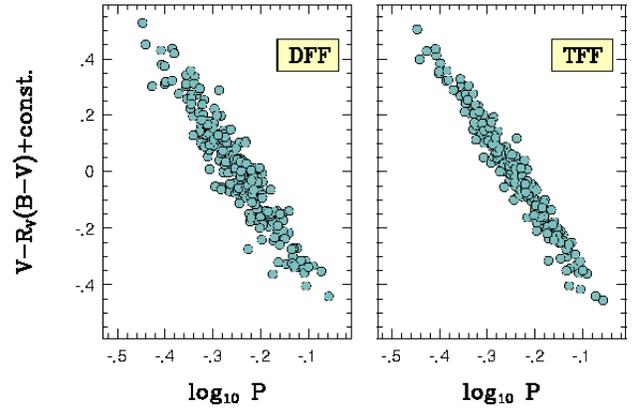}
      \caption{PLC relations computed from tests on the $248$ stars 
               of the basic dataset. Test light curves were generated 
	       with $30$ data points and additive Gaussian noise of 
	       $0.03$~mag standard deviation. The selective absorption 
	       coefficient $R_V$ is set equal to $3.1$.} 
         \label{fig10}
   \end{figure}

The test utilizes the $248$ stars of the basic dataset. For each 
star we generate test light curves in the following way:
\begin{itemize}
\item
By using the Fourier decompositions of the $V$ light curves, 
compute magnitude-averaged $V$ and $B$ colors from Eqs. (5) and 
(6) of KW01.
\item
Compute zero-averaged synthetic light curves from the Fourier 
decompositions and add the averages determined above to obtain 
noiseless light curves, with averages that satisfy exactly the 
empirical PLC relation. 
\item
Add Gaussian noise to the above noiseless light curves. The 
noise realizations used for the light curves with $V$ averages 
are different from the ones used for the light curves with the 
$B$ averages (however, they have the same standard deviation 
$\sigma$).
\end{itemize} 
We note that the above generation of test light curves is not 
entirely consistent, because the noiseless $B$ light curves 
still have Fourier decompositions corresponding to the $V$ 
light curves. However, this inconsistency has only a small 
effect on the estimated average magnitudes as we have shown 
in Sect.~5 on a smaller set of real $B$ light curves. 

In Fig.~10 we show an example of the improvement obtained by 
the application of TFF. The standard deviations around the 
best fitting straight lines are $0.064$ and $0.038$ for the 
DFF and TFF results, respectively. The derived slopes with 
their $1\sigma$ standard deviations of the means are the 
following: $-2.485\pm0.022$~(DFF) and $-2.490\pm0.013$~(TFF). 
These slopes are within the error limit of the empirical value 
of $-2.467$. Other realizations with higher noise also show 
the advantage of using TFF over DFF.

%
%

\section{Tests on the OGLE LMC data}
To test the applicability of TFF on real astronomical time series, 
we choose the RR~Lyrae database of OGLE on LMC (Soszynski et al. 
2003). Because the database contains more than 7000 entries and it 
is out of the scope of this paper to perform tests on all these 
stars, we choose a subset of it. This subset comprises fields \#~4, 5 
and 6 in the central, well-populated region of the LMC bar and 
fields \# 17, 18, 19 and 20 in the far less-populated outermost 
part. Since the observational strategy of OGLE preferred the I 
(Cousins) band, the number of data points are different in the 
various wavebands. The overall number of data points, TFF fitting 
accuracy, QF and the total number of objects are (20, 0.07, 30, 1765), 
(40, 0.06, 35, 1886) and (500, 0.06, 78, 1947) in B, V and I colors, 
respectively. We note that the fitting accuracy and QF have substantial 
star-to-star scatter. 

%
%
   \begin{figure}[h]
   \centering
   \includegraphics[width=85mm]{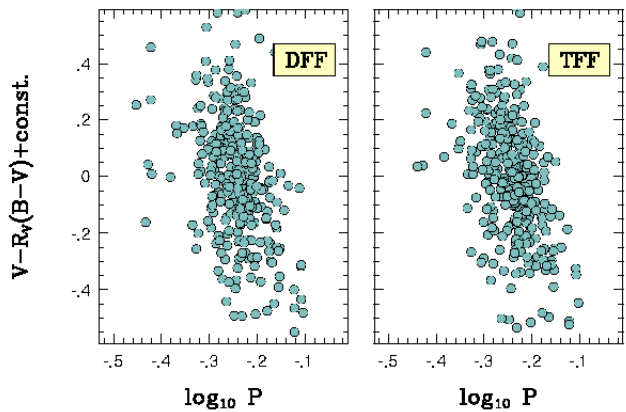}
      \caption{PLC relations computed from the DFF and TFF analyses 
               of a subset of the OGLE database of RRab stars in 
	       the LMC. The selective absorption coefficient $R_V$ 
	       is set equal to $3.1$.} 
         \label{fig11}
   \end{figure}

Two tests are performed. First we check how well the PLC 
relations of KW01 can be recovered, then we investigate the 
variation of the Fourier phase of the V light curves with the 
period. Because some light curves suffer from excessive noise 
(or some other types of deformation) we apply a parameter filter 
to keep only the good/reasonable quality light curves. In sorting out 
objects for the check of the PLC relation in (B,V) we require: 
${\rm QF}>20.0$, $\sigma_{\rm fit}/\sqrt{N}<0.02$, $A_1>0.18$, 
$0.3<(B-V)_0<0.4$. Here, except for the color index, all criteria 
refer only to the V light curves. The dereddened color difference, 
$(B-V)_0$, is computed from the observed one with the assumption 
of $E_{\rm B-V}=0.1$ (see, e.g. Kov\'acs 2000). There remained 
$318$ and $334$ stars passing these criteria in the DFF and TFF 
analyses, respectively. 

%
%
   \begin{figure}[h]
   \centering
   \includegraphics[width=70mm]{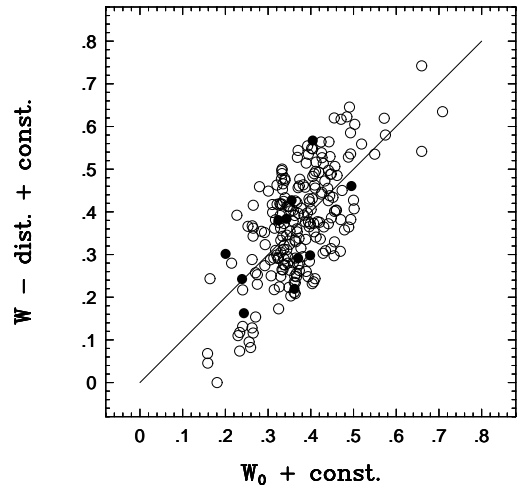}
      \caption{Correlation between $W_0=-1.59\log P$ and 
               $W=V-3.1(B-V)$ (adjusted to field averages) for 
	       218 RRab stars of selected LMC OGLE fields. The standard 
	       deviation of the fit is $0.096$~mag, the $1\sigma$ 
	       error of the slope is $0.08$. The $45\degr$ 
	       line is shown for reference. The result have been 
	       obtained by DFF with the top 100 most deviating 
	       stars clipped. Filled circles denote variables in 
	       field \#20.} 
         \label{fig12}
   \end{figure}
%

%
%
   \begin{figure}[h]
   \centering
   \includegraphics[width=70mm]{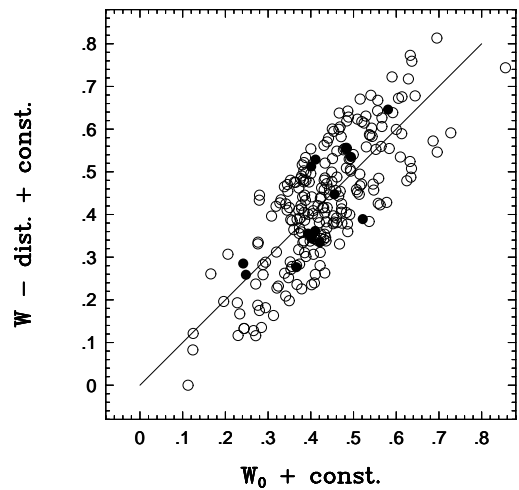}
      \caption{Correlation between $W_0=-2.33\log P$ and 
               $W=V-3.1(B-V)$ (adjusted to field averages) for 
	       234 RRab stars of selected LMC OGLE fields. The 
	       standard deviation of the fit is $0.092$~mag, 
	       the $1\sigma$ error of the slope is $0.05$. The 
	       $45\degr$ line is shown for reference. The result 
	       have been obtained by TFF with the top 100 most 
	       deviating stars clipped. Filled circles denote 
	       variables in field \#20.} 
         \label{fig13}
   \end{figure}

Figure~11 shows the resulting $\log P\rightarrow W_{\rm B-V}$ 
relations for these stars. We see that there is no particular 
improvement in the tightness of the relation whether using TFF 
or DFF. This is partially understandable, because, as Table~2 
shows, we expect only moderate improvements in both colors.  
Although this improvement is too small to be easily visible in 
the above plot, we can test the difference by employing a direct 
search for the best single-parameter regression. To take into 
account the possible zero point differences between the different 
fields, we consider each field as a different `cluster' and 
employ optimum (field-dependent) zero point shifts together with 
the uniform (i.e., field-independent) linear regression. We note 
that the zero point shifts were in general in the range of 
$\pm 0.05$~mag. 

%
%
   \begin{figure}[h]
   \centering
   \includegraphics[width=70mm]{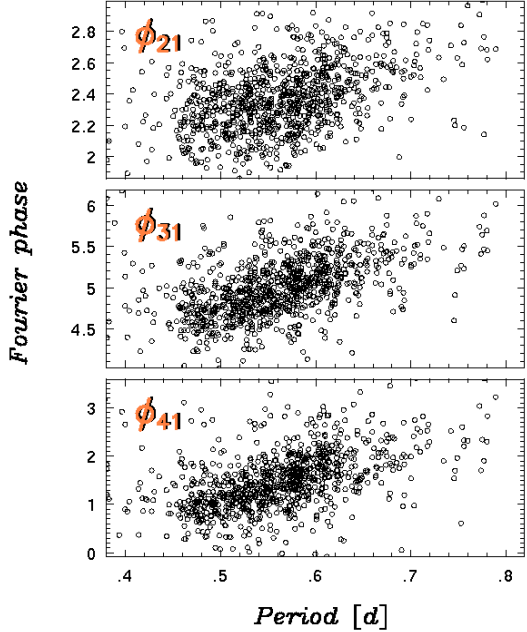}
      \caption{Low-order Fourier phases computed by DFF from the 
               V light curves of 975 RRab stars of selected LMC 
	       OGLE fields. All stars have ${\rm QF}>20$ and $A_1<0.18$.} 
         \label{fig14}
   \end{figure}
%

%
%
   \begin{figure}[h]
   \centering
   \includegraphics[width=70mm]{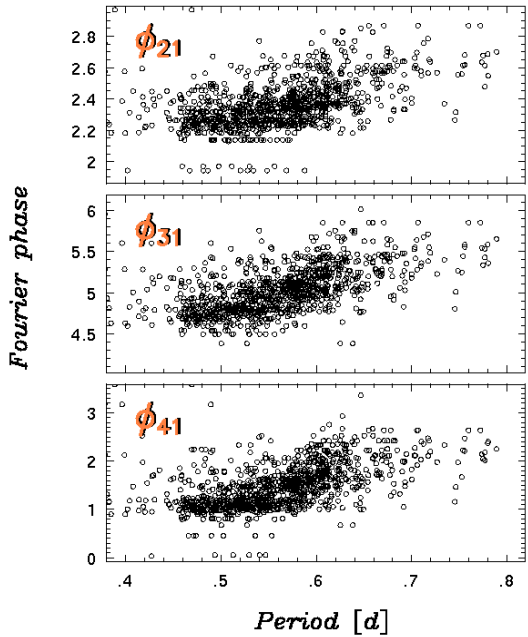}
      \caption{Low-order Fourier phases computed by TFF from the 
               V light curves of 1056 RRab stars of selected LMC 
	       OGLE fields. All stars have ${\rm QF}>20$ and $A_1<0.18$.} 
         \label{fig15}
   \end{figure}

Figures~12 and 13 show the resulting regressions. We used an iterative 
procedure in which at each step of the iteration we discarded the most 
deviating star from the best regression. We repeated this procedure 100 
times. Of course, this excessive outlier selection is not justified 
statistically. However, our goal is to see which method is capable of  
estimating the functional dependence more accurately. 

We found that in both cases the data quality was good enough for the 
code to select $P$ as the best regression parameter. This parameter 
was always the best among the other Fourier parameters at each stage 
of the iteration. However, as it is shown in the figure captions, 
TFF converged to a more accurate value of the slope of the 
$\log P\rightarrow W_{\rm B-V}$ relation. The slope obtained by DFF 
is completely out of range of the expected value of $-2.47$ (see KW01). 
It is also clear that the formal error substantially underestimates  
the true error in this case. The next best fitting Fourier parameter 
yielded standard deviations of $0.11$ and $0.13$~mag for the DFF and 
TFF methods, respectively. If we compare these values with the 
dispersions obtained for the regressions with $P$, we get that the 
increase is a factor of 1.16 for DFF, whereas it is 1.37 for TFF. 
This yields a higher significance for the correlation with $P$.        

It is worthwhile to mention that Soszynski et al. (2003) did 
not derive the $\log P\rightarrow W_{\rm B-V}$ relation. However, 
they did it for the (V, I) colors, probably because of the higher 
accuracy of the data in these colors. We note that from the 
subset used in this paper we also derived consistent slopes 
for the $\log P\rightarrow W_{\rm V-I}$ relation. We got 
$-2.584\pm 0.032$ and $-2.662\pm 0.031$ from the DFF and TFF 
analyses, respectively. The two methods perform similarly also 
in other aspects, but the TFF sample contains some 40 more stars, 
due to the better quality of the TFF fits.  

In both pairs of colors, the resulting regressions display 
larger dispersions than expected from standard statistical 
estimates. Although there might be several sources of the 
excessive scatter, crowding should definitely play a role 
(see Kiss \& Bedding 2005). Our lower limit set for the $A_1$ 
amplitude is aimed at filtering out some of the blends in a 
crude way. Obviously, a more sophisticated method is needed 
to be more successful in filtering out blended variables. 
 
A different test can be performed on the Fourier phases. This 
test is less stringent than the one presented above, because 
the quality of the result is judged from the tightness of the 
phase progression with the period, that is a good criterion 
only if there are reasonable pieces of evidence that the 
metallicity does not have a large scatter. This assumption is 
probably not a bad one for the LMC (e.g., Gratton et al. 2004 
from spectroscopy of RRab stars; Kov\'acs 2001, from double-mode 
stars). In Figs.~14 and 15 we show the progressions obtained 
for the V light curves of the selected fields mentioned at the 
beginning of this section. We see that the TFF data indeed exhibit 
a tighter correlation. Furthermore, the original number of stars 
of 1886 reduces in a less extent for TFF when various QF cutoff 
values are used. We also observe some horizontal structures in 
the TFF plots (see the sequences of circles at $\varphi_{21}=1.95$ 
and $2.15$). When lower QF cutoff is used, then these (and also 
some additional) structures become more visible. This indicates 
that the effect is partially due to poor data quality. However, 
visual inspection of the light curves producing these structures, 
and a similar test performed on the more substantial data in I 
color have shown that there are groups of stars producing nearly 
constant TFF phases even when large QF cutoff is used. Furthermore, 
because of the basic template set contains only 248 stars, it is 
possible that certain special structures observed by OGLE can be 
reproduced only by a few stars with very similar Fourier phases 
(which may yield constant phases if the template polynomial degree   
$M$ is equal to $0$ or $1$ -- a common case for poor data quality).

%
%

\section{Conclusion}
We devised a full-fetched Template Fourier Fitting (TFF) method 
for the computation of the Fourier decompositions of undersampled, 
noisy fundamental mode RR~Lyrae (RRab) light curves. The method 
can be extended to other types of variables, assuming that the 
corresponding template set is reasonably complete (i.e. it contains 
most of the `flavors' of the potential targets). The main features 
of the method are as follows.
\begin{itemize}
\item
Finds matches between the target and individual template members 
(i.e., it does not employ multi-template regressions);
\item
Fits templates to the target by applying polynomial transformation 
of the template;
\item
Optimizes the degree of the polynomial transformation between zero 
and two, depending on the data quality.
\end{itemize}
We performed a number of tests to investigate the range of applicability 
of the method. These tests included the estimation of: (i) the photometric 
iron abundance from the $\varphi_{31}$ Fourier phase and period $P$; 
(ii) the average magnitudes in various color bands; (iii) the 
period-luminosity-color relation. In all these tests TFF proved to 
perform better than an optimized Direct Fourier Fitting (DFF) method 
when the noise level was high or the number of data points was small. 
The two methods yield the same solution for light curves of high 
Signal-to-Noise Ratio (SNR). For example, when estimating [Fe/H] from 
light curves with $30$ data points we expect a factor of two increase 
in the accuracy of [Fe/H] if the standard deviation $\sigma$ of the 
noise is lower than $0.03$. For higher noise levels, TFF remains 
to be more accurate up to fairly high number of data points (e.g. with 
$\sigma=0.06$ we get better result with TFF even at $N\sim 100$).   

When applied to a subset of the OGLE RR~Lyrae database on the 
LMC (see Soszynski et al. 2003), TFF produces statistically more 
significant period-luminosity-color relation for (B, V) colors, 
although the significance of the relation still remains marginal 
on this subset. For (V, I) colors we get the same, statistically 
significant relation from both DFF and TFF. Nevertheless, the 
dispersion of the relations are high in all colors, suggesting 
the importance of crowding effects.  

The stable performance of TFF for undersampled and noisy light 
curves makes it suitable to revisit problems such as the RR~Lyrae 
metallicities in globular clusters and in galaxies (e.g. in the 
Magellanic Clouds) or the determination of average colors and 
empirical relations. With the various large-scale surveys 
(microlensing, variability, transit, etc.) there is an increase 
in the number of the good- and bad-quality light curves alike. 
Therefore, we expect TFF a useful supplementary method to the 
traditional Fourier fitting.

\begin{acknowledgements}
The support of OTKA grant K-60750 is acknowledged.
\end{acknowledgements}



\begin{thebibliography}{}
\bibitem{berto} Bertone, E., Buzzoni, A., Ch\'avez, M. \& 
                Rodr\'\i guez-Merino, L. H. 2004, AJ, 128, 829
\bibitem{clema} Clementini, G., Gratton, R. G., Bragaglia, A., 
                Ripepi, V., Fiorenzano, A. F. M., Held, E. V. \& 
		Carretta, E. 2005a, ApJ, 630, 145
\bibitem{clemb} Clementini, G., Ripepi, V., Bragaglia, A., 
                Fiorenzano, A. F. M., Held, E. V. \& Gratton, R. G. 
		2005b, MNRAS, 363, 734 
\bibitem{colli} Collister, A. A. \& Lahav, O. 2004, PASP, 116, 345 
\bibitem{dicri} Di~Criscienzo, M., Marconi, M. \& Caputo, F. 
                2004, ApJ, 612, 1092 
\bibitem{dicke} Dickens, R. J. \& Saunders, J. 1965, Roy. Obs. Bull., 
                No. 101 
\bibitem{gratt} Gratton, R. G.; Bragaglia, A., Clementini, G., 
                Carretta, E., Di Fabrizio, L., Maio, M. \& Taribello, E. 
	        2004, A\&A, 421, 937
\bibitem{jones} Jones, R. V., Carney, B. W. \& Fulbright, J. P. 
                1996, PASP, 108, 877
\bibitem{jurc2} Jurcsik, J., Benk\H o, J. M. \& Szeidl, B., 2002,
                A\&A, 390, 133
\bibitem{jurc3} Jurcsik, J. \& Kov\'acs, G. 1996, A\&A, 312, 111 (JK96)
\bibitem{kanbu} Kanbur, S. M. \& Mariani, H. 2004, MNRAS, 355, 1361
\bibitem{kissl} Kiss, L. L. \& Bedding, T. R. 2005, MNRAS, 358, 883
\bibitem{kova0} Kov\'acs, G. \& Jurcsik, J. 1997, A\&A, 322, 218
\bibitem{kova1} Kov\'acs, G. 2000, A\&A, 363, L1
\bibitem{kova2} Kov\'acs, G. 2001, A\&A, 375, 469
\bibitem{kova3} Kov\'acs, G. \& Walker, A. R. 2001, A\&A, 371, 579 (KW01)
\bibitem{kova4} Kov\'acs, G. 2003, MNRAS, 342, L58
\bibitem{kova5} Kov\'acs, G. 2005, A\&A, 438, 227
\bibitem{kunde} Kunder, A. M., Chaboyer, B., Popowski, P., Nikolaev, S., 
                Cook, K. H. 2006, PASPC, 349, 273
\bibitem{layd1} Layden, A. C. 1998, AJ, 115, 193
\bibitem{ngeow} Ngeow, C-C., Kanbur, S. M., Nikolaev, S., Tanvir, N. R.  
                \& Hendry, M. A. 2003, ApJ, 586, 959 
\bibitem{padma} Padmanabhan, N., Budav\'ari, T., Schlegel, D. J. et al. 
                2005, MNRAS, 359, 237
\bibitem{press} Press, W. H., Teukolsky, S. A., Vetterling, W. T., \& 
                Flannery, B. P., 1992, Numerical Recipes, 2nd edn., 
		Cambridge Univ. Press, Cambridge, p. 292.
\bibitem{sands} Sandstrom, K., Pilachowski, C. A. \& Saha, A. 
                2001, AJ, 122, 3212
\bibitem{simon} Simon, N. R. \& Lee, A. S. 1981, ApJ, 248, 291
\bibitem{solli} Sollima, A., Borissova, J., Catelan, M., Smith, H. A., 
                Minniti, D., Cacciari, C., \& Ferraro, F. R. 2006, 
		ApJ, 640, L43
\bibitem{soszy} Soszynski, I., Udalski, A., Szymanski, M., Kubiak, M., 
                Pietrzynski, G., Wozniak, P., Zebrun, K., Szewczyk, O. 
		\& Wyrzykowski, L. 2003, Acta Astron., 53, 93 
\bibitem{tanvi} Tanvir, N. R., Hendry, M. A., Watkins, A., Kanbur, S. M.,  
                Berdnikov, L. N. \& Ngeow, C. C. 2005, MNRAS, 363, 749
\bibitem{wolf1} Wolf, C., Meisenheimer, K. \& R\"oser, H. J. 2001, 
                A\&A, 365, 660
\end{thebibliography}
\end{document}